\documentclass[10pt, conference]{IEEEtran}
\IEEEoverridecommandlockouts
\usepackage[noadjust]{cite}
\usepackage{amsmath, amssymb, amsfonts, url}
\usepackage[ruled,vlined]{algorithm2e}
\usepackage{graphicx, tabularx, booktabs, subcaption}
\usepackage{textcomp}
\usepackage{xcolor}
\usepackage[acronym]{glossaries}

\DeclareMathOperator*{\argmin}{arg\min}
\def\BibTeX{{\rm B\kern-.05em{\sc i\kern-.025em b}\kern-.08em
    T\kern-.1667em\lower.7ex\hbox{E}\kern-.125emX}}

\begin{document}

\title{Exponential Noise Robustness of Type-Based Multiple Access for Over-the-Air Computation
\thanks{This work is part of the project SOFIA PID2023-147305OB-C32 funded by MICIU/AEI/10.13039/501100011033 and FEDER/UE.}
}

\author{\IEEEauthorblockN{Marc Martinez-Gost\IEEEauthorrefmark{1}, Ana Pérez-Neira\IEEEauthorrefmark{1}\IEEEauthorrefmark{2}\IEEEauthorrefmark{3}, 
Miguel Ángel Lagunas\IEEEauthorrefmark{2}}
\IEEEauthorblockA{
\IEEEauthorrefmark{1}Centre Tecnològic de Telecomunicacions de Catalunya, Spain\\
\IEEEauthorrefmark{2}Dept. of Signal Theory and Communications, Universitat Politècnica de Catalunya, Spain\\
\IEEEauthorrefmark{3}ICREA Acadèmia, Spain\\
\{mmartinez, aperez, malagunas\}@cttc.es
}}

\newacronym{AI}{AI}{Artificial Intelligence}
\newacronym{AirComp}{AirComp}{over-the-air computation}
\newacronym{AWGN}{AWGN}{additive white Gaussian noise}
\newacronym{CNN}{CNN}{convolutional neural network}
\newacronym{CSI}{CSI}{channel state information}
\newacronym{DA}{DA}{direct aggregation}
\newacronym{DSB}{DSB}{double sideband}
\newacronym{FL}{FEEL}{federated learning}
\newacronym{FSK}{FSK}{frequency shift keying}
\newacronym{MSE}{MSE}{mean squared error}
\newacronym{NMSE}{NMSE}{normalized mean squared error}
\newacronym{PAM}{PAM}{pulse amplitude modulation}
\newacronym{PPM}{PPM}{pulse position modulation}
\newacronym{TBMA}{TBMA}{type-based multiple access}
\newacronym{SNR}{SNR}{signal-to-noise ratio}

\maketitle
\begin{abstract}
This paper studies the robustness of type-based multiple access (TBMA) in over-the-air computation (AirComp) under nonparametric estimation, where no prior knowledge of the data distribution is available. While conventional AirComp approaches rely on amplitude modulations and suffer from noise sensitivity, TBMA enables the use of more structured modulation formats that can be exploited for improved performance. We show that the superposition of transmitted signals in TBMA induces a discrete lattice structure in the received signal space, where each lattice point corresponds to the number of devices accessing a given channel resource. By exploiting this structure through nearest-lattice-point projection, noise effects can be substantially suppressed. The proposed technique achieves an exponential decay of the mean squared error (MSE) with respect to the energy-to-noise spectral density ratio, whereas in conventional techniques the MSE only scales inversely with this ratio. Simulation results validate the theoretical findings and demonstrate that TBMA provides a fundamental robustness advantage over traditional AirComp.
\end{abstract}

\section{Introduction}

Over-the-air computation (AirComp or OAC) is an emerging communication paradigm that leverages the superposition property of wireless channels to directly compute functions of distributed data \cite{perez2024waveforms}. Rather than treating interference as a detrimental effect, AirComp exploits it as a computational resource. This approach is particularly appealing in modern wireless environments that demand the handling of massive data volumes to support real-time processing and decision-making. 

In contrast to alternative highly efficient multiple access schemes like non-orthogonal multiple access (NOMA), which are designed to reliably recover individual transmitted messages, AirComp shifts the objective from individual data reconstruction to data aggregation. More generally, the receiver is interested in computing a specific function of the distributed data, without necessarily decoding each individual contribution. This paradigm shift calls for a fundamental redesign of communication frameworks that are explicitly computation-aware. In AirComp systems, classical objectives such as minimizing the bit error rate are no longer aligned with the end goal, and new performance metrics must be considered, such as minimizing the mean squared error (MSE) or cross-entropy of the desired function, reducing computation latency, and improving energy efficiency at the devices.

Since AirComp is inherently a multipoint-to-point communication setting, it can be naturally interpreted as a multiple access problem, where the key design question lies in how users share the channel. The traditional approach to enable AirComp is known as Direct Aggregation (DA), where all users simultaneously share the same communication resource so that their signals naturally combine over the air. DA requires data to be encoded using amplitude-based transmissions, such that the superposition property of the wireless medium directly yields the desired aggregation at the receiver. By leveraging a single shared resource in this manner, DA-based AirComp has been extensively applied across a wide range of domains, including federated learning, distributed sensing and wireless data fusion \cite{perez2024waveforms}. However, this reliance on amplitude modulation makes the system particularly sensitive to noise and channel impairments, motivating the exploration of alternative modulation formats.

In contrast, angular (i.e., frequency or phase) modulations emerge as an appealing candidate due to its highest robustness. This raises a fundamental question: \textit{can users encode the information using angular-modulated signals while still enabling AirComp?} A direct extension of DA is not feasible in this case, since angular modulated signals do not combine linearly at the receiver. Nevertheless, an alternative multiple access method exists that enables the use of nonlinear modulations, commonly referred to as type-based multiple access (TBMA) \cite{mergen2006tbma}. By assigning channel resources (e.g., subcarriers) according to the data of each user, superposition occurs only among users sharing identical data values, effectively recreating an amplitude aggregation over those specific channel uses. In this way, rather than producing a single aggregated value, TBMA enables the receiver to reconstruct a histogram-like representation of the transmitted data. 

The seminal work \cite{mergen2006tbma} elaborates a pseudo-maximum likelihood estimator and establishes that TBMA is asymptotically optimal in the regime of a large number of sensors operating over noisy channels. Building on this foundation, \cite{sat-iot} leverages the Long Range (LoRa) modulation, highlighting that angular modulations can effectively benefit from AirComp in practical distributed scenarios. More recently, \cite{tbma_fl} investigates the application of TBMA to federated learning, where no prior information of the data distribution can be assumed and the original parametric estimator cannot be implemented. Alternatively, the function is computed from the received signals in a nonparametric fashion. The authors empirically show that TBMA exhibits the same MSE scaling law with respect to \gls{SNR} as DA, while achieving improved robustness to noise when the objective is to maximize the accuracy of the learned model. TBMA has been further exploited to enhance robustness against channel impairments in federated learning \cite{shain_mv}, as well as to mitigate the impact of adversarial devices in estimation and learning schemes \cite{tbma_robust}. Despite these advances, a theoretical comparison of the performance of TBMA and DA under nonparametric estimation remains largely unexplored.

In this paper, we investigate the performance of DA and TBMA under nonparametric estimation. First, we derive analytical expressions for the MSE of both schemes and show that they exhibit the same scaling behavior with respect to the inverse of the energy-to-noise spectral density ratio. Second, we introduce a novel lattice-based processing for TBMA. Specifically, we observe that the received signals in TBMA lie on a lattice structure, where each point corresponds to the number of transmitting devices per channel resource. By projecting the noisy observations onto the nearest lattice point, the impact of noise can be significantly reduced. This novel technique fundamentally changes the performance scaling, yielding an exponential improvement in the MSE with respect to the energy-to-noise spectral density ratio. This result not only highlights a significant advantage of TBMA over conventional DA, but also reinforces a broader insight: angular modulations can provide superior robustness compared to amplitude-based schemes, consistent with classical communication theory.


This paper focuses on coherent multiple access schemes, assuming perfect synchronization in time and phase among transmitters, as well as ideal channel equalization. The extension to more practical scenarios, including frequency-selective fading and synchronization impairments, is left for future work. Nevertheless, analyzing these schemes under coherent conditions is essential, as it establishes a fundamental performance baseline and provides key insights for the subsequent design and implementation of AirComp systems in realistic communication environments.

The remainder of the paper is organized as follows: Section II introduces the system model for coherent AirComp and Section III evaluates the performance of DA. Section IV presents TBMA, including both the naive aggregation scheme and the proposed lattice-aware denoising approach. Section V provides simulation results that validate and complement the theoretical analysis, and Section VI concludes the paper.



\section{System Model}
Consider a wireless network with $K$ distributed devices (i.e., transmitters) and a single server (i.e., receiver). Each device $k$ holds local data $s_k$, which could represent sensor readings or local computational results. 
We assume that the $s_k$ are independent and identically distributed (i.i.d) and take values in the discrete set $\{0, 1,\dots,N-1\}$. These values correspond to the output of an analog-to-digital converter (ADC), which is standard in practical devices. Under this assumption, quantization effects are inherently captured, allowing us to focus exclusively on performance limitations arising from the estimation of the multiple access scheme.

The goal of the network is to compute a function of the data $f(s_1,\dots,s_K)$. We focus on the sample mean function,
\begin{equation}
    f(s_1,\dots,s_K)=\frac{1}{K}\sum_{k=1}^K s_k,
    \label{eq:mean_function}
\end{equation}
the most common computation found in AirComp systems. 

Each device has access to $N$ orthogonal communication resources (e.g., time slots, frequency subcarriers, codes, etc) and define $x_{k,n}$ as the transmitted signal by device $k$ at resource $n$. Without loss of generality we assume $x_{k,n}$ to have a duration $T$ and it is sampled at $1/T$ at the receiver side. Therefore, under perfect synchronization in time and phase between all devices the received signal at resource $n$ after matched filtering is
\begin{equation}
    y_n=\sum_{k=1}^K h_{k,n}x_{k,n}+w_n
    \label{eq:rx_general}
\end{equation}
where $h_{k,n}$ is the channel coefficient between transmitter $k$ and the receiver at resource $n$ and $w_n$ is \gls{AWGN} with power $\sigma_w^2$. We can express the noise power as $\sigma_w^2=N_0B_x$, where $N_0$ is the spectral density of the power and $B_x=1/T$ is the bandwidth of the transmitted signal.

\begin{figure}[t]
    \centering
    \includegraphics[width=1\columnwidth]{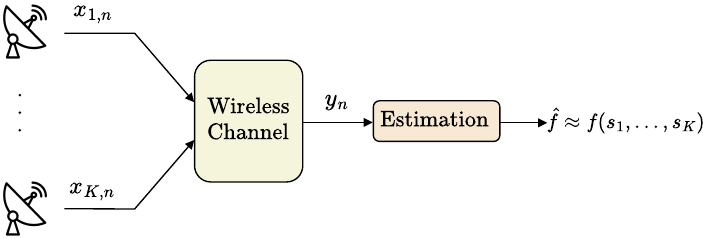}
  \caption{Baseband AirComp system model. The \textit{Estimation} block aggregates $N$ channel uses to estimate a function from the distributed data.}
  \label{fig:system_model}
\end{figure}

To better understand the capabilities of different multiple access schemes, we adopt idealized assumptions that isolate and highlight their robustness to noise in AirComp systems. Specifically, we assume perfect channel state information (CSI) at all transmitters, so that each channel coefficient can be perfectly compensated. This condition can be achieved, for example, using channel inversion methods \cite{channel_inversion}. Then, \eqref{eq:rx_general} simplifies to
\begin{equation}
    y_n=\sum_{k=1}^K x_{k,n}+w_n
    \label{eq:rx_awgn}
\end{equation}
and the transmission is only affected by noise. Figure \ref{fig:system_model} illustrates the baseband system model to compute a function from distributed data with AirComp.

To ensure a fair comparison between DA and TBMA, we evaluate the schemes under the same statistical conditions, this is, that all use $N$ channel accesses and with the same total energy $E$.

The performance metric considered is the \gls{MSE},
\begin{equation}
\text{MSE} = \mathbb{E}\big\{|f(s_1,\dots,s_K)-\hat{f}(s_1,\dots,s_K)|^2\big\},
\label{eq:figure_merit_mse}
\end{equation}
where the expectation is taken with respect to the noise.

\section{Direct Aggregation (DA)}
\label{sec:da}


The traditional approach for AirComp encodes information in the carrier’s amplitude, which can be implemented using \gls{DSB} or $M$-\gls{PAM} modulation. Since DA achieves AirComp with $N=1$, while our framework requires $N$ channel uses, we adopt a straightforward repetition strategy: DA is performed independently $N$ times, with the total available energy uniformly distributed across these $N$ transmissions. Thus, each user transmits
\begin{equation}
    x_{k,n}=\sqrt{\frac{P}{NE_S}}s_k,
    \label{eq:da_tx}
\end{equation}
where $P$ is the transmission power and $E_S$ is the average energy of $s_k$, that depends on the data distribution. Considering temporal radio resources, each with a duration $T$, the average energy per user is
\begin{equation}
    E=NT\,\mathbb{E}\left\{|x_{k,n}|^2
    \right\}=
    \frac{NTP}{NE_S}\mathbb{E}\left\{|x_{k,n}|^2
    \right\}=TP
\end{equation}

This formulation ensures that, across time, all users contribute with the same average energy. Notice that this formulation also holds for frequency radio resources. The only difference is a trade-off between the bandwidth and latency. 

The signal at the input of the receiver is
\begin{equation}
    y_n=\sum_{k=1}^K x_{k,n}+w_n=
    \sqrt{\frac{P}{NE_S}}\sum_{k=1}^K s_k+w_n,
    \label{eq:rx_da}
\end{equation}
which is a sufficient statistic of the sample mean, can be estimated as 
\begin{equation}
    \hat{f}=\frac{1}{K}\sqrt{\frac{E_S}{NP}}\sum_{n=0}^{N-1}y_n\sim\mathcal{N}\left(
    f,\frac{E_S\sigma_w^2}{K^2P}
    \right),
\end{equation}
and it is the minimum variance unbiased estimator. The corresponding MSE is
\begin{equation}
    \text{MSE}=\frac{E_S\sigma_w^2}{K^2P}=\frac{E_S}{K^2}\frac{N_0}{E},
    \label{eq:mse_da}
\end{equation}
which is inverse to the energy-to-noise spectral density ratio. Notice that DA does not experience an improvement with the number of channel uses $N$ because the energy must be equally spread over them.

\section{Type based multiple access (TBMA)}
\label{sec:tbma}

In TBMA there is a one-to-one mapping between the data space and $N$ orthogonal radio resources. Since we study the robustness to noise, without loss of generality we assume a uniform identity mapping $n=s_k$. The baseband signal transmitted by device $k$ at resource $n$ is
\begin{equation}
    x_{k,n}=
    \begin{cases}
    \sqrt{P} &
    s_k=n\\
    0 &
    s_k\neq n,
    \end{cases}
    \label{eq:tbma_tx}
\end{equation}
which can be implemented with $M$-ary \gls{FSK} or \gls{PPM}. In these two examples there is a one-to-one mapping between data and subcarriers or time shifts, respectively. Thus, in TBMA radio resources are allocated according to data, not users, enabling superposition when users have the same measurement.

For the sake of comparison with DA, we consider TBMA implemented with $N$ orthogonal time slots. Given that each user transmits in a single slot, the average energy per transmitter is
\begin{equation}
    E=NT\,\mathbb{E}\left\{|x_{k,n}|^2
    \right\}=
    NT\frac{P}{N}=TP,
\end{equation}
which corresponds to the same energy consumption of DA with $N$ repetitions in time. The key difference lies in the transmitted power: the power consumption in \eqref{eq:tbma_tx} is $N$ times smaller than in \eqref{eq:da_tx}. More importantly, in TBMA the transmit power is independent of the data, which simplifies power control considerably.

The received signal at resource $n$ is
\begin{equation}
    y_n=\sum_{k=1}^K\sqrt{P}\ \mathbb{I}(s_k=n)+w_n=
    \sqrt{P}K_n+w_n,
\end{equation}
where $\mathbb{I}(\cdot)$ is the indicator function, $K_n$ is the number of devices transmitting at resource $n$ and $w_n$ has the same distribution as $w$. Therefore, the received signals $\{y_0,\dots,y_{N-1}\}$ can be interpreted as a noisy unnormalized histogram (i.e., type) of the used resources. Due to the injective nature between $s_k$ and $n$, the received signals can be also interpreted as a nosy histogram distribution of the data.

\begin{figure*}[t]
    \centering
    \includegraphics[width=0.83\textwidth]{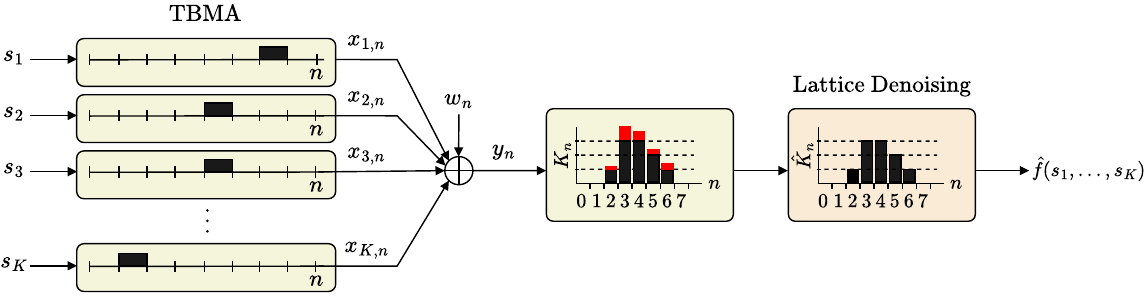}
  \caption{TBMA scheme at the transmitter side and lattice-aware denoising at the receiver side. Red bars in the first histogram illustrate the noise distortion.}
  \label{fig:tbma_robust}
\end{figure*}

\subsection{Naive Aggregation}
For any parameter of the distribution (e.g., the mean), there exists an estimator that, in the asymptotic regime $K\rightarrow\infty$, approaches the maximum likelihood estimator \cite{mergen2006tbma}. However, this approach requires a parametric model of the underlying distribution, which might be unknown apriori. This situation commonly arises in distributed sensing and learning environments, where the statistical nature of the data is typically unknown. 

Alternatively, many of the functions of interest can be computed directly by aggregating the $N$ received signals. In \cite{tbma_fl} it is shown that the sample mean can be estimated as the first moment of the received histogram:
\begin{align}
    \hat{f}=\frac{1}{\sqrt{P}K}\sum_{n=0}^{N-1} 
    n y_n&=\frac{1}{K}\sum_{n=0}^{N-1}nK_n+
    \frac{1}{\sqrt{P}K}\sum_{n=0}^{N-1}nw_n\nonumber\\
    &=\frac{1}{K}\sum_{k=1}^{K}s_k+
    \tilde{w},
    \label{eq:tbma_estimator}
\end{align}
where the first term corresponds to the sample mean $f$ and $\tilde{w}$ is noise with power  $\sigma_{\tilde{w}}^2\approx\sigma_w^2N^3/(3K^2P)$. This approximation comes from
\begin{equation}
    \sum_{n=0}^{N-1} n^2=\frac{N(N-1)(2N-1)}{6}\approx \frac{N^3}{3}
\end{equation}

Finally, the performance of TBMA with simple aggregation is
\begin{equation}
    \text{MSE}\approx \frac{N^3}{3}\frac{\sigma_w^2}{K^2P}=
    \frac{N^3}{3K^2}\frac{N_0}{E}
    \label{eq:mse_tbma_raw}
\end{equation}

This nonparametric estimation scales as DA with the SNR and poorly with the number of radio resources. In particular, \eqref{eq:tbma_estimator} recovers the measurements by exploiting the identity
\begin{equation}
    \sum_{n=0}^{N-1} nK_n=\sum_{k=1}^K s_k,
\end{equation}
which weights each noise component by its index $n$. As a result, the variance of the noise terms increases before aggregation.


It is worth emphasizing that, although TBMA is initially formulated as a multiple access scheme, it ultimately behaves as a modulation strategy, since the quality of the demodulated signal depends on a joint combination of the signals received across the $N$ resources. This is in contrast to conventional multiple access techniques such as orthogonal frequency-division multiple access (OFDMA), where the objective is to recover the signal on each resource independently, with a performance equivalent to isolated transmission. In TBMA, however, the information is inherently encoded in the collective structure of the received signals, which suggests that it is more appropriate to interpret TBMA as a modulation scheme rather than purely as a multiple access method.

\subsection{Lattice-Aware Denoising}

The naive approach of aggregating the amplitudes of the orthogonal contributions scales inversely with the SNR, as in standard DA. In contrast, the information in TBMA is primarily encoded in the relative positions of the transmitted signals across multiple channel uses rather than in their amplitudes. This structure allows the receiver to mitigate noise in the reconstructed histogram. Specifically, the normalized observations,
\begin{equation}
    \tilde{y}_n=\frac{y_n}{\sqrt{P}}= K_n+z_n,
\end{equation}
where $z_n=w_n/\sqrt{P}$. Notice that $K_n$ takes values in the set of non-negative integers $\mathbb{Z}_+$, forming a one-dimensional lattice corresponding to the number of transmitting devices per resource. The noise term $z_n$ perturbs $K_n$ away from the lattice, but estimation can be significantly improved by projecting the observations back onto the lattice.

From a detection perspective, this can be interpreted as decoding a noisy $M$-PAM signal whose alphabet is $\mathbb{Z}_+$, i.e., the number of transmitting devices. In this view, each lattice point corresponds to a valid PAM symbol, and the maximum-likelihood detector reduces to minimum-distance decoding:
\begin{equation}
    \hat{K}_n=\argmin_{k\in\mathbb{Z}_+} |\tilde{y}_n-k|=\bigl \lfloor
    \tilde{y}_n
    \bigr \rfloor,
\end{equation}
which simplifies to rounding each observation to the nearest integer. Afterwards, the nonparametric estimator in \eqref{eq:tbma_estimator} can then be applied to compute the mean. Figure \ref{fig:tbma_robust} illustrates the TBMA scheme at the transmitters and the lattice-aware denoising at the receiver.

Since the behavior of the noise depends strongly on the SNR, we will analyze the performance in two separate regimes: high SNR, where rounding effectively eliminates most small errors, and low SNR, where noise dominates and rounding has little effect.

\textbf{High SNR:} At sufficiently large SNR, the noise term becomes small relative to the symbol spacing, and detection errors are dominated by nearest-neighbor confusions. In other words, the additive noise only pushes the observation to an adjacent lattice point, causing errors of the form $K_n\pm1$. In these cases, the symbol error rate (SER) at resource $n$ is
\begin{align}
    P_{e,n} &= 1-
    \text{Pr}\left(K_n-\frac{1}{2}\leq\tilde{y}_n\leq K_n+\frac{1}{2}\right)\nonumber\\
    &=1-
    \text{Pr}\left(-\frac{1}{2}\leq z_n\leq \frac{1}{2}\right)\nonumber\\
    &=2Q\left(\sqrt{\frac{P}{4\sigma_w^2}}\right)
    =2Q\left(\sqrt{\frac{E}{4N_0}}\right),
\end{align}
which corresponds to the SER of $M$-PAM with distance of 1 between adjacent symbols. Notice that the factor 2 should be omitted for $n=\{0,N-1\}$.

Then, the MSE at resource $n$ is
\begin{align}
    \text{MSE}_n=\mathbb{E}\left\{(\hat{K_n}-K_n)^2\right\}=
    (K_n\pm1-K_n)P_{e,n}=P_{e,n},
\end{align}
since an error occurs with probability $P_{e,n}$ and has a unit magnitude.

Finally, the overall MSE can be computed considering the independence of the different noise contributions $z_n$ and the per-resource error:
\begin{align}
    \text{MSE}&=\mathbb{E}\left\{\left(
    \frac{1}{K}\sum_{n=0}^{N-1}n\hat{K}_n-\frac{1}{K}\sum_{n=0}^{N-1}n{K}_n
    \right)^2\right\}\nonumber\\
    &=\mathbb{E}\left\{\left(
    \frac{1}{K}\sum_{n=0}^{N-1}n(\hat{K}_n-K_n)
    \right)^2\right\}\nonumber\\
    &=\frac{1}{K^2}
    \sum_{n=0}^{N-1}n^2\text{MSE}_n\nonumber\\
    &\approx \frac{N^3}{3K^2}
    Q\left(\sqrt{\frac{E}{4N_0}}\right)
    \label{eq:mse_high_snr}
\end{align}

Furthermore, at high SNR the $Q$-function can be approximated by an exponential function, meaning that \eqref{eq:mse_high_snr} depends on the SNR as
\begin{equation}
    \text{MSE}\sim \exp\left(-\frac{E}{8N_0}\right)
    \label{eq:exp_behavior}
\end{equation}

\textbf{Low SNR:} The errors can be approximated as
\begin{equation}
    \hat{K}_n-K_n=\bigl \lfloor
    z_n
    \bigr \rfloor\approx z_n+q_n\approx z_n,
\end{equation}
where $q_n\sim\text{Uniform}(-0.5,0.5)$ is the quantization noise from rounding, and it is negligible compared to large Gaussian noise. Therefore, the performance at low SNR can be approximated as \eqref{eq:mse_tbma_raw}, this is, the same performance as doing no denoising.\newline

\noindent
At high SNR, TBMA exploits the fact that information is encoded in the choice of transmitted signals (i.e., channel resource) rather than in amplitude, and its MSE decreases exponentially, as seen in \eqref{eq:exp_behavior}. At low SNR, noise dominates and this structure provides little benefit, so performance does not improve over standard DA. This produces a threshold effect: once the SNR exceeds a critical value, TBMA becomes significantly more accurate than DA and should be preferred due to its exponential performance gain. The prize to pay is that TBMA consumes more resources, that is, more bandwidth if it is implemented with $M$-FSK or more time if it is implemented with PPM.

\section{Simulation Results}

We consider $s_k \sim \text{Uniform}(0, N-1)$ with $N=64$ and $K=10^3$ devices, although results hold for different distributions (e.g., Gaussian and exponential). We evaluate the schemes at the same energy-to-noise spectral density ratio $E/N_0$, which we also refer to as \gls{SNR} for simplicity (i.e., for $T=1$). The practical results are obtained via Monte Carlo simulations over $10^5$ independent realizations.

\begin{figure}[t]
    \centering
    \includegraphics[width=1\columnwidth]{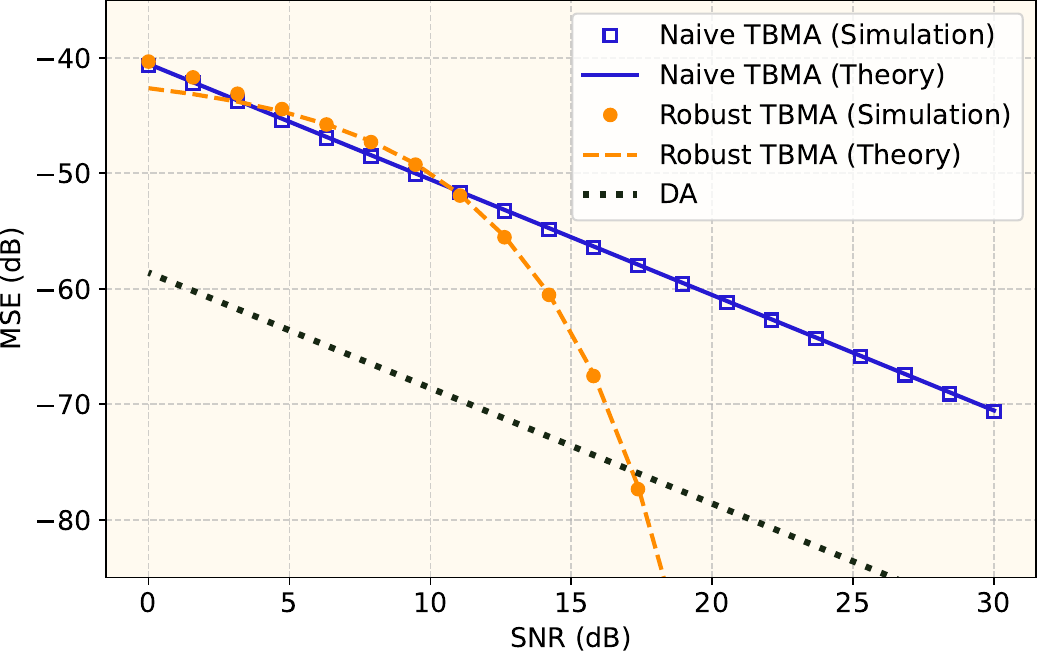}
  \caption{Theoretical and simulated MSE versus SNR for computing the sample mean with different multiple-access schemes.} 
  \label{fig:tbma_theory}
\end{figure}

Figure \ref{fig:tbma_theory} illustrates the performance of DA and TBMA, the latter with naive aggregation and lattice-aware denoising. Overall, the theoretical analysis closely matches the simulated performance, validating the accuracy of the approximations in both noise regimes. In particular, lattice-aware denoising exhibits a sharp transition around 10 dB, beyond which the MSE decreases exponentially with the SNR. Below approximately 3 dB, both TBMA schemes achieve similar performance, indicating operation in the low-SNR regime. Furthermore, TBMA outperforms DA rom around 17 dB, illustrating that TBMA is the preferred scheme in high-SNR scenarios. Likewise, this behavior indicates that TBMA achieves comparable accuracy to DA while potentially enabling significant reductions in energy consumption.

\begin{figure}[t]
    \centering
    \includegraphics[width=1\columnwidth]{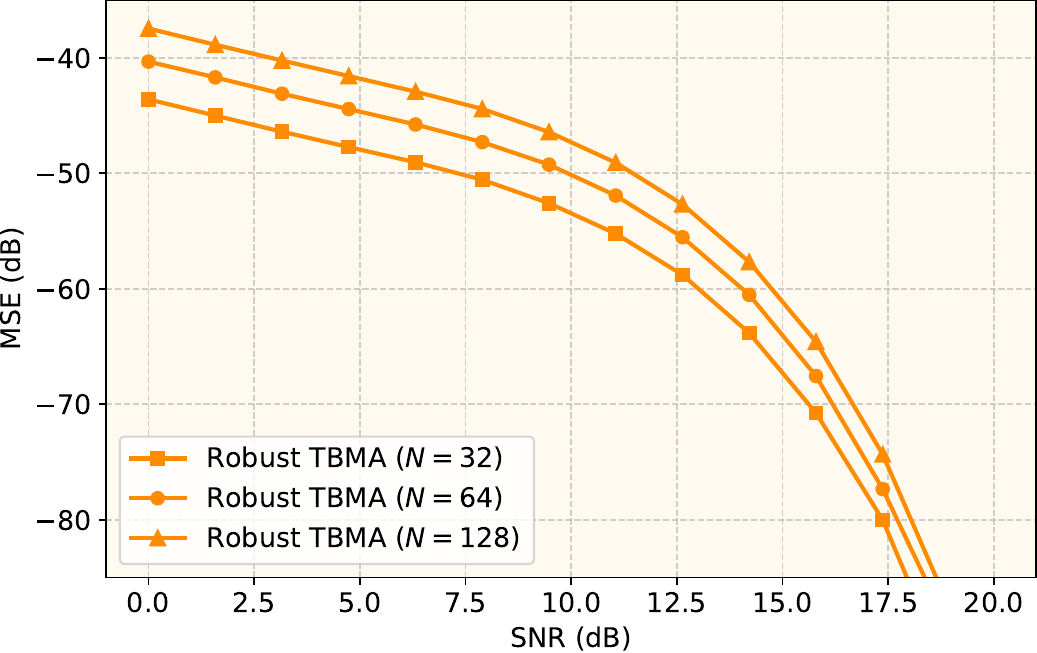}
  \caption{Performance of TBMA with lattice-aware denoising for different number of channel resources.}
  \label{fig:tbma_da}
\end{figure}

Figure \ref{fig:tbma_da} illustrates the performance of TBMA for the same uniform distribution under different numbers of channel resources $N$. It can be observed that, at a fixed SNR, the MSE consistently increases with $N$ since the error scales with $N^3$. Thus, doubling the number of resources results in an MSE increase by a factor of $10\log(2^3)\approx 9$ dB.

\section{Conclusion}

In this paper, we have shown that TBMA achieves a fundamentally different noise behavior compared to DA. The received signal in TBMA lies on a lattice, corresponding to the number of transmitting devices. This enables a denoising operation by projecting the noisy observations onto the nearest lattice point. This mechanism allows small noise perturbations to be completely removed, leading to an exponential decay of the MSE with the energy-to-noise spectral density ratio above a certain threshold. This is in contrast to DA, whose performance degrades with the inverse of the energy-to-noise spectral density ratio due to its amplitude-based nature. These gains, however, come at the cost of increased resource usage, as TBMA requires multiple channel uses to convey the data structure. This highlights a fundamental trade-off between communication resources and computation robustness in AirComp systems.

Furthermore, we have established a conceptual connection between multiple access schemes for AirComp and classical modulation techniques. In particular, the performance of DA can be interpreted as that of amplitude modulations, whereas TBMA behaves analogously to angular modulations, as both avoid encoding information in the signal amplitude and exhibit a characteristic threshold effect. This connection is further strengthened by the fact that TBMA can be effectively implemented using digital frequency modulation schemes, where information is conveyed through the selection of orthogonal frequencies rather than signal amplitude. 

As future work, it would be of interest to extend this analysis to more practical scenarios, such as fading channels and imperfect synchronization, to assess whether the robustness properties of TBMA are preserved under realistic conditions.


\bibliographystyle{IEEEbib}
\bibliography{refs}

@article{perez2024waveforms,
  title={Waveforms for Computing Over the Air},
  author={P{\'e}rez-Neira, Ana and Martinez-Gost, Marc and {\c{S}}ahin, Alphan and Razavikia, Saeed and Fischione, Carlo and Huang, Kaibin},
  journal={arXiv preprint arXiv:2405.17007},
  year={2024}
}

@ARTICLE{mergen2006tbma,
  author={Mergen, G. and Tong, L.},
  journal={IEEE Transactions on Signal Processing}, 
  title={Type based estimation over multiaccess channels}, 
  year={2006},
  volume={54},
  number={2},
  pages={613-626},
  doi={10.1109/TSP.2005.861896}}

@ARTICLE{channel_inversion,
  author={Zhu, Guangxu and Wang, Yong and Huang, Kaibin},
  journal={IEEE Transactions on Wireless Communications}, 
  title={Broadband Analog Aggregation for Low-Latency Federated Edge Learning}, 
  year={2020},
  volume={19},
  number={1},
  pages={491-506},
  doi={10.1109/TWC.2019.2946245}}

@INPROCEEDINGS{tbma_fl,
  author={Martinez-Gost, Marc and Pérez-Neira, Ana and Lagunas, Miguel Ángel},
  booktitle={GLOBECOM 2023 - 2023 IEEE Global Communications Conference}, 
  title={Frequency Modulation Aggregation for Federated Learning}, 
  year={2023},
  volume={},
  number={},
  pages={1878-1883},
  doi={10.1109/GLOBECOM54140.2023.10437413}}

@article{shain_mv,
  title={Distributed learning over a wireless network with non-coherent majority vote computation},
  author={{\c{S}}ahin, Alphan},
  journal={IEEE Transactions on Wireless Communications},
  volume={22},
  number={11},
  pages={8020--8034},
  year={2023},
  publisher={IEEE}
}

@INPROCEEDINGS{tbma_robust,
  author={Martinez-Gost, Marc and Pérez-Neira, Ana and Lagunas, Miguel Ángel},
  booktitle={2025 33rd European Signal Processing Conference (EUSIPCO)}, 
  title={Robust Over-the-Air Computation with Type-Based Multiple Access}, 
  year={2025},
  volume={},
  number={},
  pages={2042-2046},
  doi={10.23919/EUSIPCO63237.2025.11226299}}

@INPROCEEDINGS{sat-iot,
  author={Martinez-Gost, Marc and Pérez-Neira, Ana and Lagunas, Miguel Ángel},
  booktitle={2023 31st European Signal Processing Conference (EUSIPCO)}, 
  title={LoRa-Based Over-the-Air Computing for Sat-IoT}, 
  year={2023},
  volume={},
  number={},
  pages={1514-1518},
  doi={10.23919/EUSIPCO58844.2023.10289733}}

\end{document}